\documentclass[aps,showpacs,preprintnumbers,twocolumn,prl]{revtex4} 
\usepackage{graphicx,color,epsfig,rotate}
\usepackage{times,xspace}
\usepackage{amsbsy,amssymb,amsmath,bm}
\usepackage{dcolumn}
\usepackage{bm}
\usepackage{umoline}
\def\bbbc{{\mathchoice {\setbox0=\hbox{$\displaystyle\rm C$}\hbox{\hbox
to0pt{\kern0.4\wd0\vrule height0.9\ht0\hss}\box0}}
{\setbox0=\hbox{$\textstyle\rm C$}\hbox{\hbox
to0pt{\kern0.4\wd0\vrule height0.9\ht0\hss}\box0}}
{\setbox0=\hbox{$\scriptstyle\rm C$}\hbox{\hbox
to0pt{\kern0.4\wd0\vrule height0.9\ht0\hss}\box0}}
{\setbox0=\hbox{$\scriptscriptstyle\rm C$}\hbox{\hbox
to0pt{\kern0.4\wd0\vrule height0.9\ht0\hss}\box0}}}}

\begin{document}


\title{Magnetic Field Induced Transition in Vanadium Spinels}
\author{E. D. Mun$^1$, Gia-Wei Chern$^{2}$,  V. Pardo$^{3}$, F. Rivadulla$^{4}$, R. Sinclair$^{5}$, H. D. Zhou$^{5}$, V. S. Zapf$^{1}$ and C. D. Batista$^2$}
\affiliation{$^1$ NHMFL Materials Physics and Applications, $^{2}$Theoretical Division, T-4 and CNLS, Los Alamos National Lab (LANL), Los Alamos NM 87545 (USA)}%
\affiliation{$^{3}$Departamento de F{\'i}sica Aplicada and $^{4}$CIQUS, Universidad de Santiago de Compostela, 15782 Santiago de Compostela, Spain}%
\affiliation{ $^{5}$ Department of Physics and Astronomy, University of Tennessee, Knoxville, Tennessee 37996-1200 (USA)}


\date{\today}

\begin{abstract}
We study vanadium spinels $A$V$_{2}$O$_{4}$ ($A$ = Cd, Mg) in pulsed magnetic fields up to 65 T. A jump in magnetization at $\mu_{0}H$ $\approx$ 40 T is observed in the single-crystal
MgV$_{2}$O$_{4}$, indicating a field induced quantum phase transition between two distinct magnetic orders. In the multiferroic CdV$_{2}$O$_{4}$, the field-induced transition is
accompanied by a suppression of the electric polarization. By modeling the magnetic properties in the presence of strong spin-orbit coupling characteristic of vanadium spinels, we
show that both features of the field-induced transition can be successfully explained by including the effects of the local trigonal crystal field.
\end{abstract}

\pacs{75.10.Jm, 71.70.Gm, 75.50.Ee}

\maketitle

\begin{figure}
\includegraphics[width=1\linewidth]{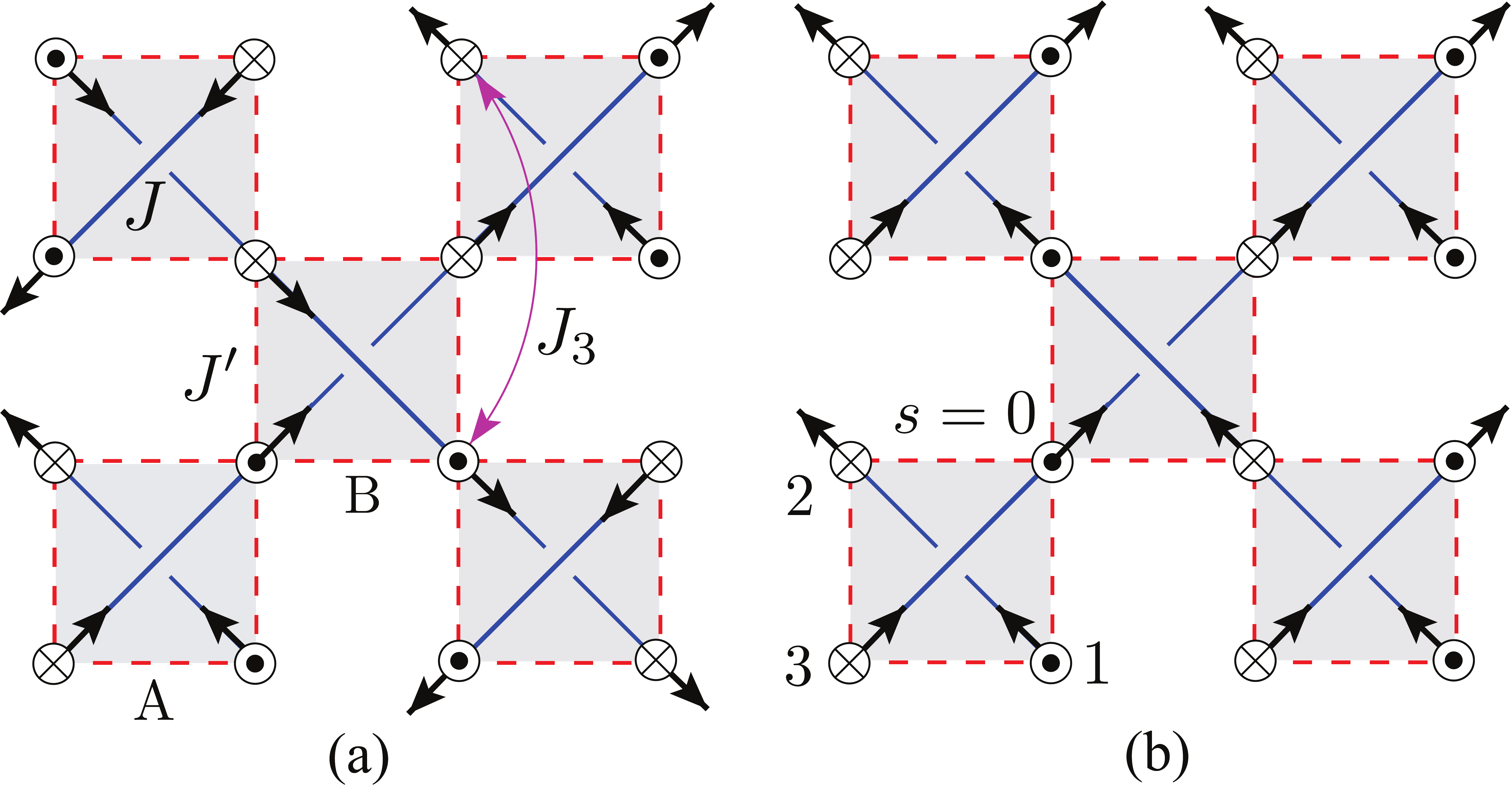}
\caption{3D orderings for AV$_2$O$_4$ viewed from $[001]$ at $H=0$ (a), and predicted to occur at high fields (b). $\odot$ and $\otimes$ denote the $+\bf z$ and
$-\bf z$ component of the moments. The arrows indicate the transverse spin components induced by the  trigonal distortion.  $J$  (solid blue)   $J'$ (dashed red) and $J_{3}$ are NN, next NN and 3rd NN exchange constants. A and B denote tetrahedra with opposite orientations. }
\label{Fig1}
\end{figure}

The interplay between spin and orbital degrees of freedom of highly frustrated magnets becomes particularly relevant when the ground state manifold of the dominant interaction is
massively degenerate \cite{Moessner06}. This is the ideal scenario for finding rich phase diagrams because small interactions become the primary selection mechanism of spin-orbital
ordering. A high susceptibility to small interactions opens the possibility of inducing phase transitions with moderate external pressure or magnetic fields. The vanadium spinels
$A$V$_{2}$O$_{4}$ ($A =$ Cd, Zn, and Mg) are archetypical realizations of highly frustrated spin-orbital systems
\cite{Ueda97,Tsunetsugu03,Tchernyshyov04,DiMatteo05,Giovannetti11,Kuntscher12,Maitra07,Reehuis03,Lee04,Vasilieva06,Bella10}. The magnetic V$^{3+}$ ions reside on a frustrated
pyrochlore lattice and contain two $d$ electrons in the three $t_{2g}$ orbitals. These materials exhibit a cubic to tetragonal transition at a temperature $T$ = $T_{S}$  and the onset
of a $\mathbf Q$ = $2\pi$(0, 0, 1) antiferromagnetic (AFM) ordering below $T_{N}$ $<$ $T_{S}$ [see Fig. \ref{Fig1}(a)] \cite{Vasilieva06,Reehuis03,Lee04,Bella10}. The pyrochlore
lattice can be viewed as a collection of cross-linking chains running along the $\langle$110$\rangle$ directions. Below $T_{N}$, the $xy$ ($z$ $\parallel$ $c$-axis) chains exhibit the
usual N{\'e}el ordering, while chains oriented along $xz$ and $yz$ directions exhibit an $\uparrow\uparrow\downarrow\downarrow$ superstructure \cite{Ueda97,Reehuis03,Lee04}. This
ordering induces an electric polarization ${\bf P}$ $\simeq$ 5$\mu{\rm C}/{\rm m}^2$ ${\hat {\bf z}}$ in CdV$_{2}$O$_{4}$, that arises from different oxygen
displacements along $xz$ and $yz$ bonds \cite{Giovannetti11}, giving opposite contributions to ${\bf P}$ depending on whether the bond is ferromagnetic (FM) or AFM. Because the {\it
magnitude} of the displacements is different for FM and AFM bonds, the $\uparrow\uparrow\downarrow\downarrow$ structure induces a net ${\bf P} \parallel {\hat {\bf z}}$.

The ground state manifold of these spinels is extensively degenerate if only nearest-neighbor (NN) exchange is included. The  $\mathbf Q = 2\pi
(0,0,1)$ ordering must then be selected by residual interactions which arise as a sequence of two selection mechanisms. The tetragonal distortion reduces the frustration by increasing  the exchange along the $xy$ chains and inducing AFM spin correlations ~\cite{Lee04}. This distortion is accompanied by 
long-range ferro-orbital ordering (occupied $xy$ orbitals) and the onset of a magnetic  easy $z$-axis. However, the exchange  between  crossing $xy$ chains remains frustrated. In the 
Mott limit, relevant for CdV$_2$O$_4$, the $\uparrow\uparrow\downarrow\downarrow$ ordering is stabilized by a weak third NN AFM exchange.~\cite{Tsunetsugu03} For spinels near the Mott transition, like ZnV$_{2}$O$_{4}$ \cite{Pardo08,Kato12}, the $\uparrow\uparrow\downarrow\downarrow$ superstructure could
originate  from Fermi point nesting of the quasi-1D $yz$ and $xz$ chains \cite{Chern11}.

The weak nature of the interactions that select the  magnetic ordering in highly frustrated magnets suggests that moderate  fields could induce a different spin
ordering. Indeed, multiple magnetic field-induced  transitions have been reported in pyrochlore antiferromagnets and spin ice $R_2$Ti$_2$O$_7$
\cite{ramirez02,higashinaka,tabata,rule06,yin13}. While most  transitions result from the competition between various spin-spin interactions and the Zeeman coupling, the presence
of orbital degrees of freedom makes vanadium spinels ideal candidates for studying similar phenomena in frustrated spin-orbital systems. By including the
subtle interplay between spin-orbit coupling and  lattice distortion, our  model predicts a new high-field magnetic order induced by a local trigonal distortion, which was not regarded as an important factor in previous studies of vanadium spinels.

Here we show experimental evidence of a quantum phase transition  induced by relatively small fields in  MgV$_{2}$O$_{4}$ and CdV$_{2}$O$_{4}$ spinels, in spite of their large
Curie-Weiss temperature $|\Theta_{{\rm CW}}|$ $\simeq$ 300 K \cite{Vasilieva06}. For a single-crystal of MgV$_{2}$O$_{4}$, we observe a clear jump in the magnetization at $\mu_0 H\simeq 40$T, while
the multiferroic ordering of powder CdV$_{2}$O$_{4}$ is suppressed for $\mu_{0}H$ $\gtrsim$ 30 T. We explain this transition with a new selection mechanism based on a
trigonal distortion, which is intrinsic to spinel structures and rotates the local easy-axis toward the $\langle 111 \rangle$ direction of each V$^{3+}$. This
rotation induces a weak FM component perpendicular to the N\'eel order parameter of each $xy$ chain [see Fig. \ref{Fig1}]. While the FM components of different chains cancel out for the
low-field ${\mathbf Q} = (001)$ structure of Fig. \ref{Fig1}(a), the ${\mathbf Q} = {\mathbf 0}$ spin ordering shown in Fig. \ref{Fig1}(b) acquires a net transverse FM
component, which makes it energetically favorable for large enough $H$. This new structure is not ferroelectric, which is in agreement with our experiments.

\begin{figure}
\includegraphics[width=1\linewidth]{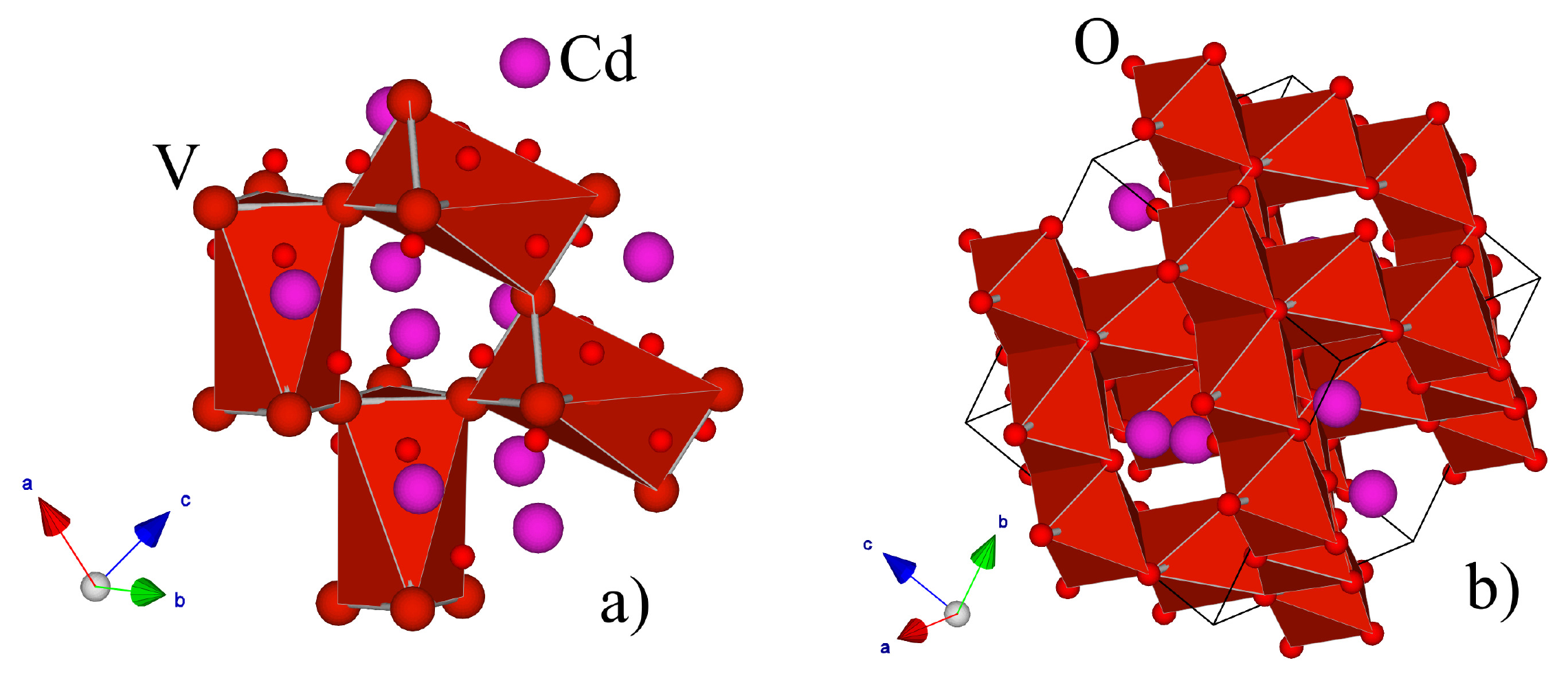}
\caption{(a) Trigonal antiprism environment of the neighboring V atoms. (b)  Octahedra  of O around the V atoms. (Pink circles with no bonds drawn are Cd$^{2+}$ ions.) The two fields act in
opposite directions,  but they do not cancel out.}%
\label{Fig2}
\end{figure}

Ref.~\cite{yaresko} shows how the electronic structure of $A$M$_2$X$_4$ spinels containing $t_{2g}$ electrons is affected by a varying trigonal field. One component arises from the
trigonal distortion of the oxygen octahedron, as shown in Fig. \ref{Fig2}. An opposing effect is caused by the surrounding trigonal anti-prism of $t_{2g}$-based metals (V ions). Both
contributions eventually cancel when moving from ZnCr$_{2}$O$_{4}$ to HgCr$_{2}$O$_{4}$ because the oxygen trigonal field increases while the Cr trigonal field decreases. For
CdV$_{2}$O$_{4}$ and the structure proposed in \cite{cdvo_struct}, a very small tetragonal field is expected below $T_S$ that is accompanied by a larger trigonal distortion. The
structure proposed in \cite{Giovannetti11} (and also in \cite{Pardo08} for the related compound ZnV$_{2}$O$_{4}$) explains the origin of the additional tetragonal term that further
stabilizes the $d_{xy}$ orbital. From {\em ab initio} calculations we get $\Delta$ = 350 meV and $\delta$ = 250 meV for the $t_{2g}$ splittings induced by the tetragonal and trigonal distortions, respectively~\cite{Pardo08}.

We first present $P(H)$ and $M(H)$ data for CdV$_{2}$O$_{4}$ and show that the multiferroic $\uparrow\uparrow\downarrow\downarrow$ ordering of CdV$_{2}$O$_{4}$ is suppressed for
$\mu_{0}H$ $>$ 30 T. Similar features in $M(H)$ are observed in single-crystalline MgV$_{2}$O$_{4}$, which is too conductive to measure $P(H)$. Polycrystalline samples of
CdV$_{2}$O$_{4}$ were prepared by solid-state reaction, and the single crystals of MgV$_{2}$O$_{4}$ by a travelling floating zone method, as described in the Supplement. Low-field $M(H)$ and 
$P(H)$ measurements are consistent with previous publications \cite{Giovannetti11,Islam12}.

The high-field data  are plotted in Fig. \ref{Fig3}, and more details are described in the Supplement.  For CdV$_{2}$O$_{4}$, $M(H)$ increases linearly until a field between 30 and 40 T and then shows an upturn for $T < T_{S}$ [see Fig. \ref{Fig3}(a)].  The $M(H)$ curve of CdV$_{2}$O$_{4}$ reaches 0.4 $\mu_{B}$/V$^{3+}$
at $\mu_{0}H$ = 65 T, which is far below the saturated value for V$^{3+}$ \textbf{S} = 1 (2$\mu_{B}$/V$^{3+}$). By linear extrapolation, the saturation value would be reached for
$\mu_{0}H$ $>$ 200 T, consistent with $|\Theta_{\rm CW}| \simeq$ 300 K.

\begin{figure}
\includegraphics[width=1\linewidth]{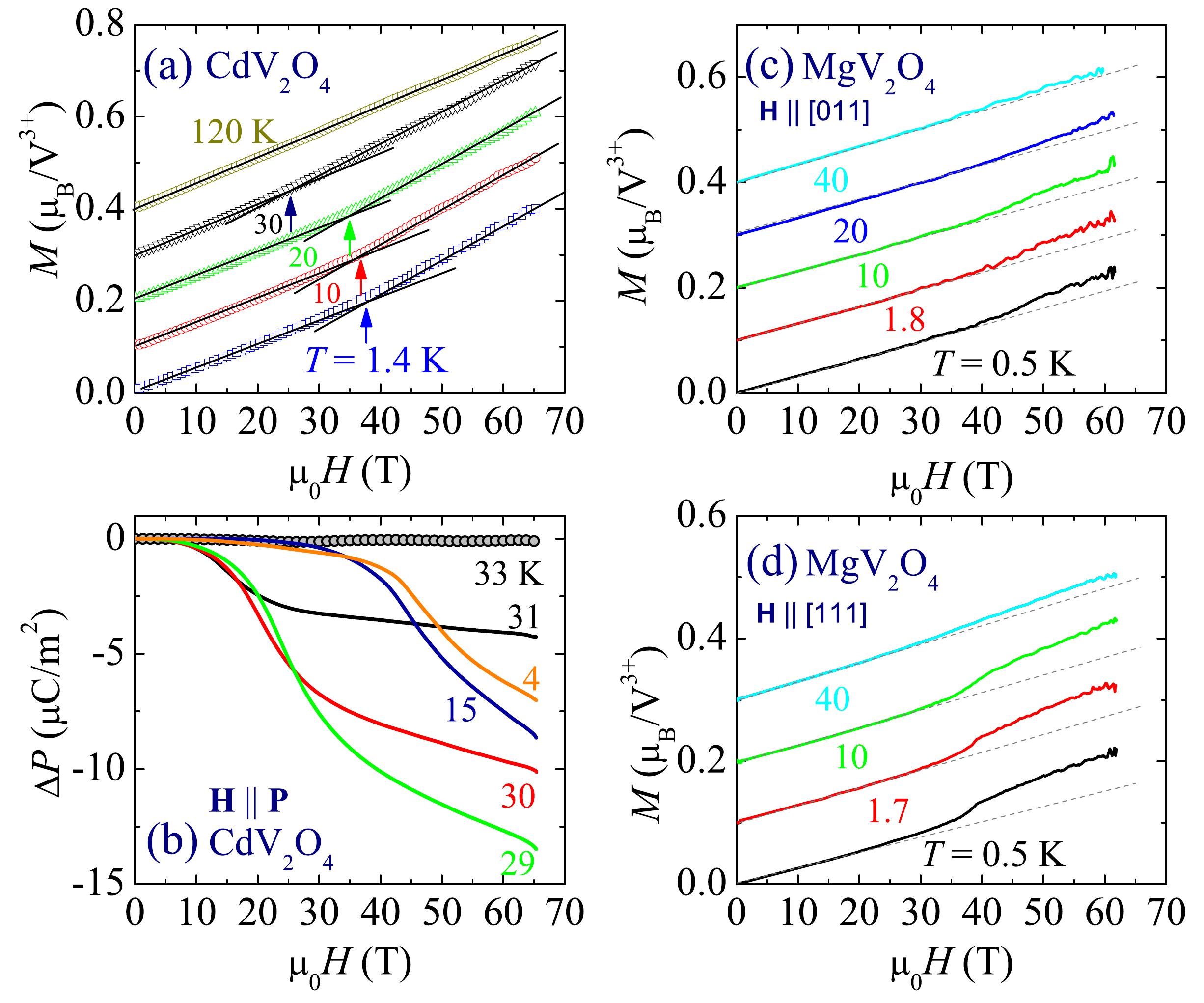}
\caption{(a) Magnetization isotherms, $M(H)$, of CdV$_{2}$O$_{4}$ at selected temperatures. Vertical arrows represent the critical field determined by drawing two straight lines. (b)
$\Delta P(H)$ at selected temperatures, determined by integrating the d$\Delta P$/d$t$ as a function of time. The sample was poled by applying external voltage across the contacts:
$E_{P}$ = 670 kV/m. (c) and (d) $M(H)$ of MgV$_{2}$O$_{4}$ at select temperatures for orientations (1) and (2) described in the Supplement. For clarity, the $M(H)$ curves are
successively shifted upward by 0.1 $\mu_{B}$/V$^{3+}$. Dashed-lines are guides to the eye.}%
\label{Fig3}%
\end{figure}%

Fig. \ref{Fig3}(b) shows $\Delta P(H)$ of CdV$_{2}$O$_{4}$ for \textbf{H} $\parallel$ \textbf{P} (similar data for \textbf{H} $\perp$ \textbf{P} are shown in the Supplement).
$P(H)$ remains constant up to a field between 20 and 40 T, depending on the temperature, and then begins to decrease. No change in $P$ with $H$ is observed for $T$ $>$ $T_{N}$ = 32.5 K (see 33 K curve). Just below $T_N$ ($T=29$K),  $P$(65 T) - $P$(0)  is close to the value of $\Delta P(T)$ across temperature-induced phase transition at $T_N$
\cite{Giovannetti11} indicating that the 65 T  field mostly suppresses ferroelectricity near $T_N$. However, $P$(65 T) - $P$(0) shrinks for $T$ $\leq$ 29 K, indicating that a finite $P$ remains for $ H > 65$ T.

Finally, Figs.~\ref{Fig3}(c) and (d) show $M(H)$ curves of MgV$_{2}$O$_{4}$ for \textbf{H} $\parallel$ [011] and \textbf{H} $\parallel$ [111], respectively. MgV$_{2}$O$_{4}$ is too electrically conductive to measure electric polarization. Like for
CdV$_{2}$O$_{4}$, $M(H)$ is linear up to a field between 30 and 40 T and then shows an upturn or a jump. In the Supplement, we describe a magnetic hysteresis observed for  \textbf{H} $\parallel$ [111].

\begin{figure}
\includegraphics[width=1\linewidth]{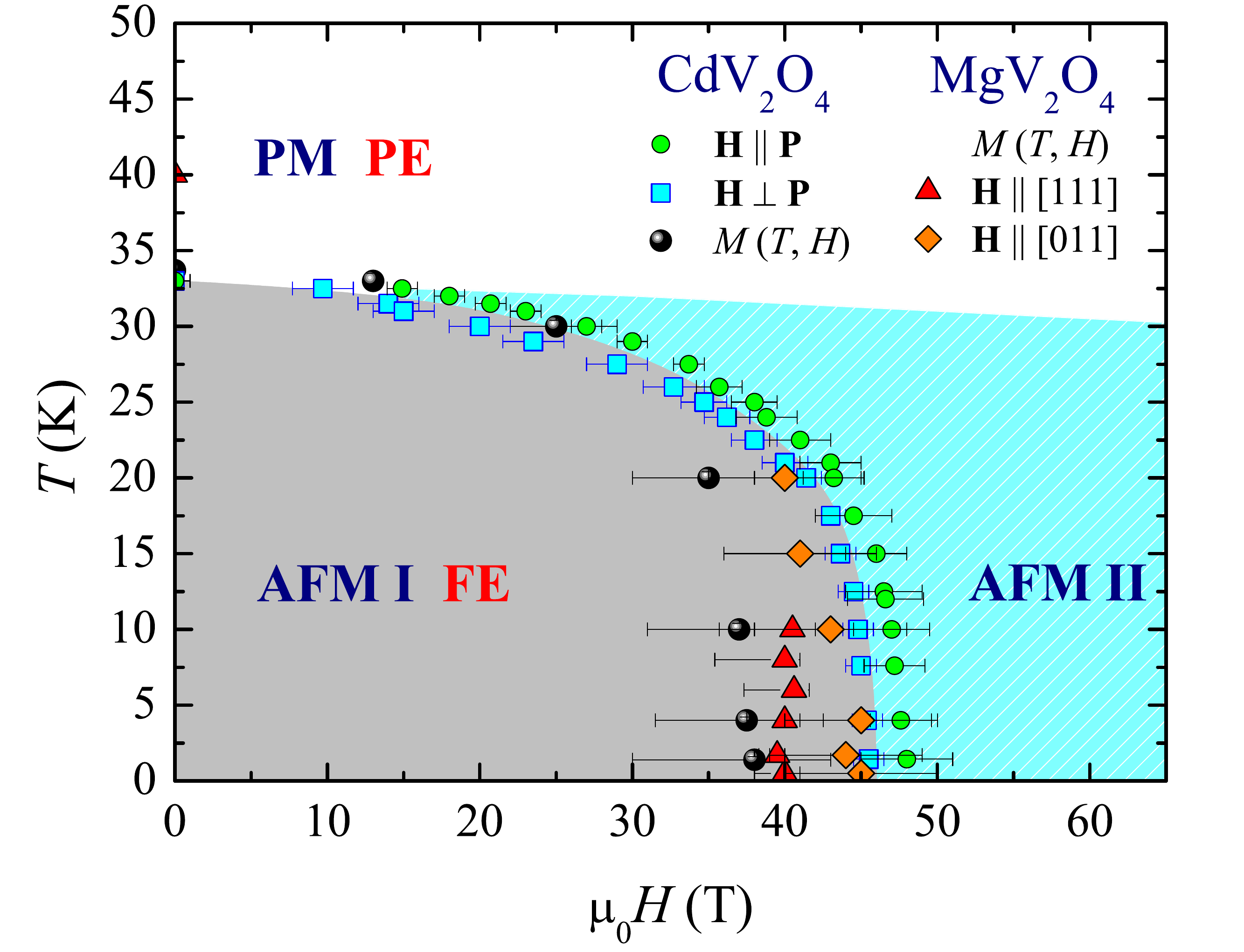}
\caption{$H-T$ phase diagrams of CdV$_{2}$O$_{4}$ and MgV$_{2}$O$_{4}$, obtained from $M(T, H)$ and  $\Delta P(H)$ measurements. Abbreviations PM,  PE, and FE are for paramagnetic, paraelectric, and ferroelectric state. The shaded area is the FE state and the lined area represents a mixed PE-FE state due to
the polycrystalline nature of CdV$_{2}$O$_{4}$. The FE phase boundary is inferred from the maximum of d$\Delta P$/d$\mu_{0}H$.}
\label{Fig4}%
\end{figure}

Fig.~\ref{Fig4} shows the $H-T$ phase diagram of CdV$_{2}$O$_{4}$ and MgV$_{2}$O$_{4}$ that results from the measured $M(H)$ and $\Delta P(H)$ curves. Because CdV$_{2}$O$_{4}$ is
polycrystalline, we find it likely that the observed behaviors of $P(H)$ and $M(H)$ are caused by a single field-induced transition that is broadened by anisotropy, {\it i.e.}, the
beginning of the upturn in $M(H)$ and downturn in $P(H)$ correspond to the transition field for ${\bf H}$ parallel to the easy-axis, while the transition field along the hard axis is
not reached by 65 T. The phase diagram includes peaks in d$\Delta P$/d$\mu_{0}H$ and the onset of the deviation from linear behavior in $M(H)$ (determined by intersecting the linear
behaviors below and above the upturn).

We now introduce a model Hamiltonian that includes the effect of the tetragonal and trigonal distortions:
\begin{eqnarray}
    \label{eq:H}
    \mathcal{H} &=& J \sum_{\langle ij \rangle}^{xy} \mathbf S_i\cdot\mathbf S_j + J' \sum_{\langle ij \rangle}^{yz, zx} \mathbf S_i\cdot\mathbf S_j
    - D \sum_i \left(\mathbf S_i \cdot \hat{\mathbf n}_i\right)^2 \nonumber \\
    & & + J_3 \sum_{\langle\langle\langle ij \rangle\rangle\rangle} \mathbf S_i \cdot\mathbf S_j  - g \mu_B \sum_i \mathbf H \cdot \mathbf S_i.
\end{eqnarray}
$J$ is the NN AFM exchange that couples the V$^{3+}$ spins along the $xy$ chains and $J'$ is the NN interaction along $xz$ and $yz$. $D$ is the single-ion anisotropy, $\hat{\mathbf
n}_i$ is a unit vector along the easy-axis, $J_3$ is the third NN exchange, $g$ is the gyromagnetic factor, and $\mu_B$ is the Bohr magneton.

As in other vanadium spinels \cite{Tsunetsugu03,DiMatteo05}, $J$ and $J'$ depend on the occupancy of the relevant $d$ orbitals. The occupancy of the $d_{xy}$-orbital, and thus the
strength of $J$, becomes dominant after the tetragonal distortion below $T_S$. $J'$ is determined by the orbital ordering of the remaining $d_{xz}$ and $d_{yz}$ orbitals. Two types of
orbital  states have been proposed for vanadium spinels: staggered real~\cite{Tsunetsugu03} and a uniform complex orbital ordering~\cite{Tchernyshyov04}. Both cases result in $|J'| <
J$. Because $J'$  is also geometrically frustrated, the final 3D ordering depends on  residual perturbations.  Here we  consider two competing perturbations: the Zeeman
coupling to the external field and a 3rd NN exchange $J_3$. The 3rd NN pairs are separated by twice the NN distance on the same $\langle 110 \rangle$ chains [Fig. \ref{Fig1}(a)].

The $D$ term originates from the relativistic spin-orbit interaction. We estimate the orientation of the easy axes $\hat{\mathbf n}_i$ and the anisotropy strength $D$ by diagonalizing
the single-ion Hamiltonian $\mathcal{H}_{t_{2g}} = \Delta \hat\lambda_8 + \delta (\hat\lambda_1 + \hat\lambda_4 + \hat\lambda_6) - \lambda \mathbf L'\cdot \mathbf S$ in the two
$d$-electron basis. $\hat\lambda_i$ are Gell-mann matrices, $\mathbf L'$ is the effective angular momentum of length $L' = 1$ for the $t_{2g}^2$ electronic configuration, and $
\lambda \approx 20$ meV is the effective spin-orbit coupling constant. The crystal-field splitting estimated from our {\em ab initio} calculations ($\Delta \approx 350$~meV and
$\delta \approx 250$~meV)  leads to $D \approx 15.5$ meV and an easy axis $\hat{\mathbf n}_i$, which is tilted about $\theta = 35^\circ$ from the $z$ axis toward the local $\langle 111
\rangle$ direction. Specifically, the easy axes at the four sublattices of the pyrochlore are $\hat{\bf n}_m = \cos\theta\, \hat{\bf z} + \sin\theta\, \hat{\bf e}_m$, where the
in-plane unit vectors $\hat{\bf e}_m$ point along the $[110]$, $[\bar 1 1 0]$, $[1\bar 1 0]$, and $[\bar 1 \bar 1 0]$ directions for $m = 0, 1, 2, 3$, respectively (see Fig.
\ref{Fig1}).  We note that the essential physics discussed below is largely independent of the specific values of these parameters.

We first consider the zero field magnetic order. The frustration of the $J'$ couplings between the crossing $xy$-chains is relieved by $J_3$. For zero trigonal
distortion, $\delta$ = 0, $J_3$ favors a collinear $\uparrow\uparrow\downarrow\downarrow$ ordering of spins along the $yz$ and $xz$ chains (the 3rd neighbor pairs on the $xy$ chains
are parallel to each other due to a dominant $J$), giving rise to the $\mathbf Q$ = 2$\pi$(0,0,1) 3D  ordering shown in Fig. \ref{Fig1}(a).
To characterize the AFM order in the pyrochlore lattice we introduce two N\'eel order parameters $\mathbf L_x =  \mathbf S_0 + \mathbf S_1 - \mathbf S_2 - \mathbf S_3 $ and $\mathbf
L_y = \mathbf S_0 + \mathbf S_2 - \mathbf S_1 - \mathbf S_3 $ for a tetrahedron~\cite{Chern06}, where $ \mathbf S_m $ is the magnetization of sublattice $m$. The corresponding values
for the  structure shown in Fig. \ref{Fig1}(a) are $\mathbf L_x(\mathbf r_{\rm A}) = 4S\, \hat{\mathbf z}\, e^{i \mathbf Q\cdot\mathbf r_{\rm A}}$ and $\mathbf L_y(\mathbf r_{\rm B})
= 4S\, \hat{\mathbf z}\, e^{i \mathbf Q\cdot\mathbf r_{\rm B}}$, where $\mathbf r_{\rm A}$ and $\mathbf r_{\rm B}$ are coordinates of tetrahedra of type-A and B, respectively. Because
$\exp\left(i \mathbf Q \cdot \mathbf r_{\rm A, B}\right) = \pm 1$, the sign of the N\'eel order parameters alternate between successive layers.

The trigonal crystal field $\delta$ rotates the easy axis and gives rise to a net magnetization $\mathbf M =  \mathbf S_0 + \mathbf S_1 + \mathbf S_2 + \mathbf S_3$ in each
tetrahedron. This small FM component is modulated: $\mathbf M(\mathbf r_{\rm A}) \approx 4 S \sin\theta\, \hat{\mathbf y} \, e^{i\mathbf Q\cdot\mathbf r_{\rm A}}$, $\mathbf M(\mathbf
r_{\rm B}) \approx 4 S \sin\theta\, \hat{\mathbf x} \, e^{i\mathbf Q\cdot\mathbf r_{\rm B}}$, and the net magnetization vanishes: $\langle \mathbf M(\mathbf r) \rangle = 0$.  The
tetragonal symmetry is preserved by this magnetic order as the system is invariant under a $\pi/2$ rotation in the $xy$ plane ( A $\leftrightarrow$ B)
accompanied by the exchange $\mathbf L_x \leftrightarrow \mathbf L_y$. Because the $\mathbf Q$ = (001) order is selected by $J_3$ out of many degenerate states consisting of decoupled AFM $xy$ chains, the Zeeman coupling to a large enough magnetic field
should overcome $J_3$ and select the state with finite  $\mathbf M$ that is shown in Fig.~\ref{Fig1}(b). We suggest that the transition between these two ordered phases explains our measurements. The magnetic order in Fig. \ref{Fig1}(b) has a wave vector ${\mathbf Q} = {\mathbf 0}$, {\it i.e.}, all
tetrahedra are in the same magnetic structure. This state is described by magnetic order parameters: $\mathbf L_x(\mathbf r) \approx 4S\,\cos\theta\,\hat{\mathbf z}$, $\mathbf
L_y(\mathbf r) = 0$, and $\mathbf M(\mathbf r) = 4S\,\sin\theta\, \hat{\mathbf y}$, for both types of tetrahedra. The tetragonal symmetry is broken  in this case and the Zeeman energy
gain is $E_Z = 4S\, g\mu_B \sin\theta\, H$ per tetrahedron for a transverse magnetic field $\mathbf H = (0, H, 0)$. A first-order quantum phase transition thus takes place at $g \mu_B
H_c \approx \mbox{const}\times J_3/\sin\theta$ that removes the polarization by suppressing the $\uparrow\uparrow\downarrow\downarrow$ structure.

\begin{figure}
\includegraphics[width=0.9\linewidth]{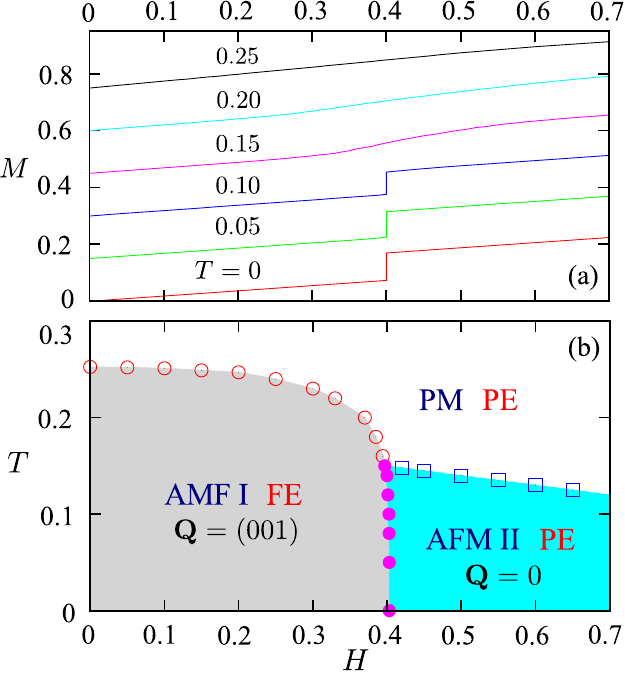}
\caption{(a) $M(H)$ along the $x$ or $y$ directions at various temperatures. The field is measured in units of $J / g\mu_B$. (b) $T$--$H$ phase diagram of model~(\ref{eq:H}) obtained
from MC simulations. $T$ is measured in units of $J$. The open circles and squares denote second-order phase boundaries, while filled circles mark a first-order transition line. The
three phase boundaries intersect at a bicritical point.}%
\label{Fig5}
\end{figure}

We verified the above picture by performing classical Monte Carlo (MC) simulations of ${\cal H}$ for $J' = 0.16J$, $J_3 = 0.01 J$, and $D = 0.7 J$. ($S = 1$ spins are approximated by
classical unit vectors). We use the standard Metropolis algorithm and periodic boundary conditions for lattices of  up to $N_s = 16\times 8^3$ spins. Fig. \ref{Fig5}(a) shows the
$M(H)$ curves for ${\bf H} \parallel {\hat {\bf y}}$  obtained at different temperatures. A sharp discontinuity at $g\mu_B H \approx 0.4 J$ and $T=0$ indicates a first-order
transition. The discontinuity  decreases with increasing temperature and disappears at $T \approx 0.15 J$. The resulting $H$--$T$ phase diagram [Fig. \ref{Fig5}(b)] includes the two ordered phases shown in Figs. \ref{Fig1}(a) and (b). The transition to the high-$T$
paramagnetic phase is always continuous (the phase boundaries were estimated by the crossing of the Binder's cumulant). The first-order line between the two  ordered phases was
determined with the method of mixed initial state  (each of the two coexisting orders occupies half of the lattice)~\cite{Creutz79}. Relating  $T_c (H=0)$
with the experimental value of $T_c \approx 33$ K (Fig. \ref{Fig4}), we estimate a transition field $\mu_0 H_c \approx$ 47 T, in good agreement with our measurements. The 
first-order transition is consistent with the magnetic hysteresis observed in  MgV$_{2}$O$_{4}$ (Supplement).

In summary, we observe a field-induced phase transition marked by a magnetization jump at $H \approx$ 40 T in a
single-crystal of MgV$_{2}$O$_{4}$, and an upturn in the $M(H)$ slope  of  polycrystalline CdV$_{2}$O$_{4}$. We also present a  model, which includes the effect of
orbital degrees of freedom, lattice distortion, and spin-orbit interactions, and predicts a field-induced $\mathbf Q$ = 0 magnetic order that is stabilized by relatively small
magnetic fields in comparison to the dominant exchange. The measured field-induced transition is attributed to the onset of the $\mathbf Q$ = 0 order. Contrary to the $\mathbf Q$ =
2$\pi$(0, 0, 1) zero field magnetic ground state, the new field-induced state does not support ferroelectricity, in agreement with the suppression of ${\bf P}$ that we observe in the
multiferroic spinel CdV$_{2}$O$_{4}$. This $\mathbf Q$ = 0 state is only possible in the presence of a small trigonal distortion, which has been observed in many vanadium spinels. For example, the trigonal distortion is essential for stabilizing the staggered orbital order of the ferrimagnetic vanadate MnV$_{2}$O$_{4}$ \cite{sarkar,chern10}. Our theory thus underscores the
importance of including the trigonal distortion, that exists in both zero and applied fields, for describing the magnetism of vanadium spinel compounds. In particular, the presence of the trigonal distortion in vanadium spinels indicates that the orbital ordering
in the vanadium spinels family is most likely of the staggered type~\cite{Tsunetsugu03}, as indeed the case for CdV$_2$O$_4$.
Finally, because the $\mathbf Q$ = 0 state also breaks the tetragonal
symmetry, a similar phase transition might also be driven by epitaxial strain.

We thank N. Perkins and D. Khomskii for useful discussions and Brian Scott for single crystal orientations. Work at the LANL was performed under the auspices of the U.S.\ DOE contract
No.~DE-AC52-06NA25396 through the LDRD program. The NHMFL facility is funded by the NSF through cooperative grant DMR-1157490,  by the DOE and by the State of Florida. This material
is based upon work supported in part by the NSF under Grant No. PHY-1066293 and the hospitality of the ACP. V.P. acknowledges financial support from the Ramon y Cajal Program.

\end{document}